\documentclass[a4paper]{jpconf}
\usepackage{graphicx}

\usepackage{fancyhdr}

\usepackage{iopams}
\usepackage{graphics,wrapfig,times}

\usepackage{cite}

\usepackage{amsfonts}
\usepackage{mathrsfs}
\usepackage{amsmath,amssymb}
\usepackage[T1]{fontenc}

\begin{document}


\title{Poincar\'e gauge theory with even and odd parity dynamic connection modes: isotropic Bianchi cosmological models}

\author{Fei-Hung Ho$^{1}$, James M. Nester$^{1,2,3}$}

\address{$^1$Department of Physics, National Central University, Chungli 320,
Taiwan\\
$^2$Institute of Astronomy, National Central University, Chungli 320, Taiwan\\
$^3$Center for Mathematics and Theoretical Physics,
National Central University, Chungli 320, Taiwan}

\ead{93242010@cc.ncu.edu.tw, nester@phy.ncu.edu.tw}

\begin{abstract}
The Poincar\'e gauge theory of gravity
 has a metric compatible connection with independent dynamics
that is reflected in the torsion and curvature. The theory allows
two good propagating spin-0 modes.
Dynamical investigations using a simple expanding cosmological model found that
the oscillation of the 0$^+$ mode could account for an accelerating expansion similar to that presently observed.
The model has been extended to include a $0^{-}$ mode and more recently cross parity couplings.
We investigate the dynamics of this model in a situation which is simple, non-trivial,
and yet may give physically interesting results that might be observable.
We consider homogeneous cosmologies, more specifically, isotropic Bianchi class A models.
We find an effective Lagrangian for our dynamical system,  a system of first order equations,
 and present some typical dynamical evolution.
\end{abstract}

\section{Introduction}


All the known fundamental physical interactions can be  formulated
in a common framework: as {\em local gauge theories}.
However the standard theory of gravity, Einstein's general relativity (GR), based on the spacetime metric, is a
rather unnatural gauge theory.
Physically (and geometrically) it is reasonable to consider gravity as a
gauge theory of the local Poincar\'e symmetry of Minkowski
spacetime.  A theory of gravity based on local spacetime geometry gauge symmetry, the quadratic Poincar\'e gauge theory of gravity (PG, aka PGT) was worked out some time ago.

We  briefly sketch this theory, and how the search for good dynamical propagating modes led to focusing on the two scalar modes, and then how cosmological models were used to reveal the dynamics of the resulting model.

There is no known fundamental reason why the gravitational coupling should respect parity. With this in mind,
the general quadratic PG theory has recently had a major extension to include all possible odd parity couplings. The appropriate odd parity couplings have been incorporated, in particular, into the special case of the dynamically favored two scalar mode model.  Here we show how a simple effective Lagrangian can reveal in a cosmological context the dynamics of this extended model.

\section{The Poincar\'e gauge theory}

In the Poincar\'e gauge theory of gravity~\cite{HHKN,Hehl80,HS80,MieE87,HHMN95,GFHF96,Blag02},
the two sets of local gauge potentials are, for ``translations'',
the orthonormal co-frame $\vartheta^\alpha=e^\alpha{}_i {\rm d}x^i$,
where the metric is
$g=-\vartheta^0\otimes\vartheta^0+\delta_{ab}\vartheta^a\otimes\vartheta^b$,
and, for ``rotations'', the metric-compatible (Lorentz Lie-algebra
valued) connection 1-forms
$\Gamma^{\alpha\beta}=\Gamma^{[\alpha\beta]}{}_i {\rm d}x^i$. The
associated field strengths are the torsion and curvature 2-forms
\begin{eqnarray}
T^\alpha&:=&{\rm d}\vartheta^\alpha+\Gamma^\alpha{}_\beta\wedge
\vartheta^\beta=\frac12 T^\alpha{}_{\mu\nu}\vartheta^\mu\wedge\vartheta^\nu,
\label{torsion}\\
R^{\alpha\beta}&:=&{\rm d}\Gamma^{\alpha\beta}
+\Gamma^\alpha{}_\gamma\wedge\Gamma^{\gamma\beta}=
\frac12R^{\alpha\beta}{}_{\mu\nu}\vartheta^\mu\wedge\vartheta^\nu,
\label{curvature}
\end{eqnarray}
which satisfy the respective Bianchi identities:
\begin{equation}
DT^\alpha\equiv R^\alpha{}_\beta\wedge \vartheta^\beta,\qquad
DR^\alpha{}_\beta\equiv0.\label{bianchi}
\end{equation}

Turning from kinematics to dynamics, the PG Lagrangian density is generally taken
to have the standard quadratic Yang-Mills form, which leads to quasi-linear second order equations for the gauge potentials. Qualitatively,
\begin{equation}
\quad{\mathscr L}[\vartheta,\Gamma]\sim \kappa^{-1}[\Lambda+\hbox{curvature}+\hbox{torsion}^2]+\varrho^{-1}\hbox{curvature}^2,
\end{equation}
where $\Lambda$ is the cosmological constant, $\kappa=8\pi G/c^4$, and $\varrho^{-1}$ has the dimensions of action.
The field equations obtained by variation w.r.t. the two gauge potentials have the respective general forms
\begin{eqnarray}
\Lambda+\hbox{curvature}+D\hbox{torsion}+\hbox{torsion}^2+\hbox{curvature}^2
&\sim&\hbox{source energy-momentum density},\\
\hbox{torsion}+D\hbox{curvature}&\sim&\hbox{source spin density}.
\end{eqnarray}
From these two equations, with the aid of the Bianchi
identities (\ref{bianchi}), one can obtain, respectively,
the conservation of source energy-momentum and angular momentum
statements.

Earlier investigations considered only models with even parity terms; the models had in general 10 dimensionless
scalar coupling constants.
The recent BHN investigation~\cite{BHN11} systematically considered
all the possible odd parity Lagrangian terms, introducing 7 new pseudoscalar coupling constants.
Not all of these coupling constants are physically independent, since there are 3 topological invariants: the (odd parity) Nieh-Yan~\cite{NY} identity
\begin{equation}
d(\vartheta^\alpha\wedge T_\alpha)\equiv T^\alpha\wedge T_\alpha+R_{\alpha\beta}\wedge\vartheta^{\alpha\beta},
\end{equation}
the (even parity) Euler 4-form $R^{\alpha\beta}\wedge R^{\gamma\delta}\eta_{\alpha\beta\gamma\delta}$,
and the (odd parity) Chern-Simons 4-form $R^\alpha{}_\beta\wedge R^\beta{}_\alpha$.

Early PG Investigations (especially~\cite{HS80,SN80}) of the linearized
theory identified six possible dynamic connection modes; they
carry spin-$2^{\pm}$, spin-$1^{\pm}$, spin-$0^{\pm}$. A good dynamic
mode should transport positive energy and should not propagate
outside the forward null cone. The linearized investigations found
that at most three modes can be simultaneously dynamic; all the
acceptable cases were tabulated; many combinations of three modes
are satisfactory to linear order. Complementing this, the
Hamiltonian analysis revealed the related constraints~\cite{BMNI83}.
Then detailed investigations of the Hamiltonian and propagation~\cite{CNY98,HNZ96,yo-nester-99,yo-nester-02} concluded that effects
due to nonlinearities in the constraints could be expected to render
all of these cases physically unacceptable except for the two
``scalar modes'', carrying spin-$0^+$ and spin-$0^-$.

In order to further investigate the dynamical possibilities of these PG scalar modes,  Friedmann-Lema\^{i}tre-Robinson-Walker (FLRW)
 cosmological models were considered.  Using a $k=0$ model it was found that the $0^+$ mode naturally couples to the acceleration of the universe and could account for the present day observations~\cite{YN07,SNY08}; this model was then extended to include the $0^-$ mode~\cite{JCAP09}.

After developing the general odd parity PG theory, in BHN~\cite{BHN11} the two scalar torsion mode PG Lagrangian was extended to include the appropriate pseudoscalar coupling constants.

The BHN Lagrangian~\cite{BHN11}
has the specific form
\begin{eqnarray}
{\mathscr L}_{\hbox{\footnotesize BHN}}[\vartheta,\Gamma]&=&\frac{1}{2\kappa}\biggl[- 2 \Lambda + a_0
R + b_0X -\frac12\sum_{\mathsf{n}=1}^3 a_{\mathsf{n}}
{{}^{\mathsf{(n)}}\!T}{}^2+ 3\sigma_2
T_\mu P^\mu\biggr] \nonumber\\
&& \qquad\qquad\ \  -\frac{1}{2\varrho}\biggl[\frac{w_6}{12}R^2-\frac{w_3}{12}X^2+\frac{\mu_3}{12}RX\biggr], \label{Ldensity}
\end{eqnarray}
where
$R$ is the scalar curvature and $X$ is the
pseudoscalar curvature (specifically $X/6=R_{[0123]}$ is the
magnitude of the one component of the totally antisymmetric
curvature), and $T_\mu\equiv T^\alpha{}_{\alpha\mu}= {}^{(2)}\!
T{}^\alpha{}_{\alpha\mu}$,
$P_\mu\equiv\frac{1}{2}\epsilon_{\mu\nu}{}^{\alpha\beta}T^\nu{}_{\alpha\beta}
=\frac{1}{2}\epsilon_{\mu\nu}{}^{\alpha\beta} {}^{(3)}\!T{}^\nu{}_{\alpha\beta}$
are the torsion trace and axial vectors.

In the aforementioned work the general field equations were worked out, and then specialized to find the most general 2-scalar mode PG FLRW cosmological model.  Here we will consider manifestly isotropic Bianchi models.

\section{The PG scalar mode isotropic Bianchi I and IX cosmological models}
For earlier PG cosmological investigations see Minkevich and coworkers, e.g.,~\cite{Min80,Min83,MN95,MG06,MGK07}
 and Goenner \& M\"uller-Hoissen~\cite{GMH};
for recent work see~\cite{BHN11,YN07,SNY08,JCAP09,LSX09a,LSX09b,ALX10,WW09}.

For the usual FLRW models, although they are actually homogeneous and isotropic, the representation is not manifestly so.  Indeed they are merely manifestly isotropic-about-a-chosen-point.  In contrast, the representation used here for the isotropic (class A) Bianchi I and IX models have the virtue of being {\em manifestly homogeneous\/} and {\em manifestly isotropic}.

For homogeneous, isotropic  Bianchi type I and IX
(respectively equivalent to FLRW $k=0$ and $k=+1$) cosmological models the
isotropic orthonormal coframe has the form
\begin{equation}
\vartheta^0:=dt,\qquad \vartheta^a:=a\sigma^a,
\end{equation}
where $a=a(t)$ is the scale factor and $\sigma^j$ depends on the (not needed here) spatial coordinates in such a way that
\begin{equation}
d\sigma^i=\zeta\epsilon^i{}_{jk}\sigma^j \wedge\sigma^k,
\end{equation}
where $\zeta=0$ for Bianchi I and $\zeta=1$ for Bianchi IX, thus $\zeta^2=k$,
the sign of the FLRW Riemannian spatial curvature.

Because of isotropy, the only non-vanishing connection one-form
coefficients are of the form
\begin{equation} \Gamma^a{}_0=\psi(t)\,
\sigma^a,\qquad \Gamma^a{}_b=\chi(t)\epsilon^a{}_{bc}\, \sigma^c.
\end{equation}
Here $\epsilon_{abc}:= \epsilon_{[abc]}$ is the usual 3 dimensional
Levi-Civita anti-symmetric symbol.

From the definition of the
curvature (\ref{curvature}), one can now find all the nonvanishing
curvature 2-form components:
\begin{eqnarray}
R^a{}_b&=&\dot \chi dt\wedge
\epsilon^a{}_{bc}\sigma^c+[\psi^2-\chi^2]\sigma^a\wedge \sigma_b
+ \chi\zeta\epsilon^a{}_{bc}\epsilon^c{}_{ij}\sigma^i\wedge\sigma^j,\\
R^a{}_0&=&\dot \psi dt\wedge \sigma^a
-\chi\psi \sigma^b\wedge\epsilon^a{}_{bc}\sigma^c
+\psi\zeta\epsilon^a{}_{bc}\sigma^b\wedge\sigma^c
.
\end{eqnarray}
Consequently, the scalar and pseudoscalar curvatures are,
respectively,
\begin{eqnarray}
R&=&6[a^{-1}\dot \psi+a^{-2}(\psi^2-[\chi-\zeta]^2+\zeta^2)], \label{R} \\
X&=&6[a^{-1}\dot \chi+2a^{-2}\psi(\chi-\zeta)].\label{X}
\end{eqnarray}

Because of isotropy, the only nonvanishing  torsion tensor
components are of the form
\begin{equation}
T^a{}_{b0}=u(t)\delta^a_b, \qquad T^a{}_{bc}=-
2x(t)\epsilon^a{}_{bc},
\end{equation}
where $u$ and $x$ are the scalar and pseudoscalar torsion, respectively.
From the definition of the torsion (\ref{torsion}), one can find the
relation between the torsion components and the gauge variables:
\begin{equation}
u=a^{-1}(\dot a-\psi), \qquad x= a^{-1}(\chi-\zeta). \label{fchi}
\end{equation}
Note that $a^{-1}\psi=H-u$, where $H=a^{-1}\dot a$ is the {\em Hubble function}.

From the isotropic assumption, the material energy momentum tensor
must have the perfect fluid form.  In our present work we focus on the late
time behavior. Accordingly we assume that the fluid pressure can be
neglected, so that the gravitating material behaves like dust with a
density satisfying $\rho a^3=\hbox{constant}$. Also, although we expect the spin density to play an
important role in the early universe, it is reasonable to assume
that the material spin density is negligible at late times.

\section{Effective Lagrangian}

The dynamical equations for these homogeneous cosmologies could be obtained by imposing the Bianchi symmetry on the general field equations found in~\cite{BHN11} from their {\em Lagrangian density}.  On the other hand dynamical equations can (following the same procedure used in~\cite{JCAP09}) be obtained directly and independently from a classical mechanics type effective {\em Lagrangian}, which in this case can be simply obtained by restricting the BHN Lagrangian density to the Bianchi symmetry (this step is where the manifestly homogeneous representation plays an essential role).  This procedure is known to successful give the correct dynamical equations for all Bianchi class A models (which includes our cases) in GR (see~\cite{AS91}), and it is conjectured to also work equally as well for the PG theory.  Our calculations explicity verify this property for the isotropic Bianchi I and IX models.  Indeed the equations we obtained in this way are equivalent to those found (at a later date) by BHN for their FLRW models (which are equivalent to our isotropic Bianchi models) by restricting to FLRW symmetry their general dynamical PG equations. This has proved to be a useful cross check.

The {\em effective Lagrangian}
is
\begin{eqnarray}
L_\mathrm{eff}= (2\kappa)^{-1}(a_0R+b_0X-2\Lambda)a^3
&+&\frac3{2\kappa}\bigl(-a_2u^2+4a_3x^2+4\sigma_2ux\bigr)a^3\nonumber\\
&-&\frac1{24\varrho}\bigl(w_6R^2-w_3X^2+\mu_3RX\bigr)a^3.
\end{eqnarray}
It should be noted that the parameter restrictions $a_2<0$, $w_6<0$, $w_3>0$, and $\mu_3^2+4w_3w_6<0$
are {\em necessary} for the {\em least action principle}, which requires {\em positive\/} quadratic-kinetic-terms.

In the following we often take for simplicity units such that $\kappa=1=\varrho$.  These factors can be easily restored in the final results by noting that in the Lagrangian they occur in conjunction with certain PG parameters.  Hence in the final results one need merely make the replacements
\begin{eqnarray}
\{a_0,a_2,a_3,b_0,\Lambda,\sigma_2\}\to \kappa^{-1}\{a_0,a_2,a_3,b_0,\Lambda,\sigma_2\}, \quad\{w_3,w_6,\mu_3\}\to \varrho^{-1}\{w_3,w_6,\mu_3\}.
\end{eqnarray}

The effective gravitational Lagrangian has the usual form of a sum of terms homogeneous in ``velocities'': $L_\mathrm{eff}=L_0+L_1+L_2$; the associated {\em energy function} is thus
\begin{eqnarray}
{\cal E}_\mathrm{eff}&:=&\frac{\partial L_\mathrm{eff}}{\partial \dot \psi}\dot\psi+\frac{\partial L_\mathrm{eff}}{\partial \dot\chi}\dot\chi+\frac{\partial L_\mathrm{eff}}{\partial \dot a}\dot a-L_\mathrm{eff}=L_2-L_0\nonumber\\
&=&a^3\Biggl\{-3(a_0-\frac12a_2)u^2-3a_0H^2+3x^2(a_0-2a_3)\nonumber\\
&&\quad\ \ +6uH(a_0-\frac12a_2)+6(b_0+\sigma_2)x(H-u)-3a_0\frac{\zeta^2}{a^2}+\Lambda\nonumber\\
&&\quad\ \ -\frac{w_6}{24}\left[R^2-12R\left\{(H-u)^2-x^2+\frac{\zeta^2}{a^2}\right\}\right]\nonumber\\
&&\quad\ \ +\frac{w_3}{24}\left[X^2+24Xx(H-u)\right]\nonumber\\
&&\quad\ \ -\frac{\mu_3}{24}\left[RX-6X\left\{(H-u)^2-x^2+\frac{\zeta^2}{a^2}\right\}+12Rx(H-u)\right]\Biggr\}.\label{energyfunction}
\end{eqnarray}
Since $L_{\rm eff}$ is time independent, the energy function is conserved.
This constant {\em energy} has the form $-a^3\rho$ where $\rho$ can be identified as the material {\em energy density}. This relation is consistent with the vanishing divergence of the source fluid energy-momentum tensor, which has been assumed to have the dust form with vanishing pressure.

Making use of the formulas for the torsion and curvature components
in terms of the gauge variables (\ref{R},\ref{X},\ref{fchi}), we now
obtain the Euler-Lagrange equations.
The dynamical equations are the $\psi$ equation: 
\begin{eqnarray}
\frac{d}{dt}\frac{\partial L_\mathrm{eff}}{\partial \dot\psi}
   &=&\frac{\partial L_\mathrm{eff}}{\partial \psi},\qquad \Longrightarrow\nonumber
\end{eqnarray}
\begin{equation}
-\frac{w_6}2\dot R-\frac{\mu_3}4 \dot
X=-\left[3(2a_0-a_2)-w_6R-\frac{\mu_3}2
X\right]u-\left[6(b_0+\sigma_2)-\frac{\mu_3}2 R+w_3X\right]x. \label{Epsi}
\end{equation}
%
%
%
%
 the $\chi$ equation:
\begin{eqnarray}
\frac{d}{dt}\frac{\partial L_\mathrm{eff}}{\partial \dot\chi}
    =\frac{\partial L_\mathrm{eff}}{\partial \chi},\qquad\Longrightarrow\nonumber
\end{eqnarray}
\begin{equation}
-\frac{\mu_3}4 \dot
R+\frac{w_3}2\dot X=-\left[6(b_0+\sigma_2)-\frac{\mu_3}2 R+w_3X\right]u+\left[6(a_0-2a_3)-w_6R-\frac{\mu_3}2
X\right]x, \label{Echi}
\end{equation}
%
%
 and the $a$ equation:
\begin{eqnarray}
\frac{d}{dt}\frac{\partial L_\mathrm{eff}}{\partial \dot a}
   =\frac{\partial L_\mathrm{eff}}{\partial a}
    ,\qquad\Longrightarrow\nonumber
\end{eqnarray}
\begin{eqnarray}
-3(a_2\dot u-2\sigma_2 \dot x)
 &=&(a_0R+b_0X-3\Lambda)+6H(a_2u-2\sigma_2x)\nonumber\\
 &&+\frac3{2}(-a_2u^2+4a_3x^2+4\sigma_2ux)
   -\frac1{24}(w_6R^2-w_3X^2+\mu_3RX)\nonumber\\
 &&+\frac{a_0}2\left[-6(H-u)^2+6x^2-6\frac{\zeta^2}{a^2}\right]+6b_0x(H-u)\nonumber\\
 &&-\frac1{24}\left(2w_6R+\mu_3X\right)\left[-6(H-u)^2+6x^2
   -6\frac{\zeta^2}{a^2}\right]\nonumber\\
 &&+\frac1{24}\left(2w_3X-\mu_3R\right)\left[12x(H-u)\right]. \label{Ea}
\end{eqnarray}


\bigskip
The above equations (\ref{Epsi},\ref{Echi},\ref{Ea}) are 3 {\em second order} equations for the gauge potentials $a,\psi,\chi$.
However they can in an alternative way be used as part of a set of 6 {\em first order} equations along with the Hubble relation
$
\dot a=aH
$ 
and the following two relations, obtained by taking the time
derivatives of the torsion (\ref{fchi}) and using the curvature definitions (\ref{R},\ref{X}):
\begin{eqnarray}
\qquad\dot x&=&-Hx-\frac{X}6-2x(H-u),\\
\dot H -\dot u&=&\frac{W}6-H(H-u)-(H-u)^2+x^2-\frac{\zeta^2}{a^2}.
\end{eqnarray}
One advantage of such a reformulation is that the variables are now all observables.

Our 6 first order dynamical equations and the energy constraint equation can now be put in the form
\begin{eqnarray}
\qquad\qquad\qquad\dot a&=&aH,\label{dota}\\
\qquad\qquad\qquad\dot H&=&
  \frac{1}{6a_2}( \tilde a_2R-2\tilde \sigma_2 X)-2H^2
  +\frac{\tilde a_2-4\tilde a_3}{a_2}x^2-\frac{\zeta^2}{a^2}
+\frac{\rho}{3a_2}+\frac{4\Lambda}{3a_2}\label{dotH},\\
\qquad\qquad\qquad\dot u&=&
  -\frac1{3a_2}(a_0R+\tilde \sigma_2 X)-3Hu
  +u^2-\frac{4a_3}{a_2}x^2
  +\frac{\rho}{3a_2}+\frac{4\Lambda}{3a_2}\label{dotu}, \\
\qquad\qquad\qquad\dot x&=&-\frac{X}6-(3H-2u)x\label{dotx},\\
\quad-\frac{w_6}2\dot R-\frac{\mu_3}4 \dot X&=&
  \left[3\tilde a_2+w_6R+\frac{\mu_3}2
X\right]u+\left[-6\tilde \sigma_2+\frac{\mu_3}2 R-{w_3}X\right]x\label{dotRdotX}\\
\quad\,\,\,-\frac{\mu_3}4 \dot R+\frac{w_3}2\dot X &=&
  \left[-6\tilde \sigma_2+\frac{\mu_3}2 R-w_3X\right]u
  -\left[12\tilde a_3+{w_6}R+\frac{\mu_3}2X\right]x\label{dotXdotR},
\end{eqnarray}
\begin{eqnarray}
\rho&=&
         3(-\frac12 \tilde a_2+2\tilde a_3)x^2+\frac{3a_2}{2}\left[H^2
       +\frac{\zeta^2}{a^2}\right]-\Lambda\nonumber\\
  &&-\left(6\tilde \sigma_2+w_3X-\frac{\mu_3R}{2}\right)x(H-u)
       +\frac{1}{24}(w_6R^2-w_3X^2+\mu_3RX)\nonumber\\
  &&-\left(\frac32 \tilde a_2+\frac{w_6R}{2}+\frac{\mu_3X}{4}\right)
       \left[(H-u)^2-x^2+\frac{\zeta^2}{a^2}\right], \label{rho}
\end{eqnarray}
where we have introduced certain {\em modified} parameters
$\tilde{a}_2$, $\tilde{a}_3$ and $\tilde{\sigma}_2$
with the definitions
\begin{equation}\label{mp}
    \tilde{a}_2:=a_2-2a_0, \qquad\tilde{a}_3:=a_3-\frac{a_0}{2}, \qquad\tilde{\sigma}_2:=\sigma_2+b_0.
\end{equation}
The last two dynamical equations (\ref{dotRdotX},\ref{dotXdotR}) could of course easily be resolved for $\dot R$ and $\dot X$, but for some puposes the form given above is more convenient.

It is remarkable that here in these effective Lagrange equations and the
associated conserved energy we get (for this case with dust as the source)
exactly the correct equations for our model---{\em without including
any explicit source coupling.}

From the above first order equations (following the method used in~\cite{JCAP09}) we have found the associated linear equations and the late time normal modes.  The details of these results will be reported in another work.  In the following we will show some typical evolution.


\subsection{Numerical Demonstration}
\begin{figure}[thbp]
\begin{tabular}{rl}
\includegraphics[width=0.333\textwidth]{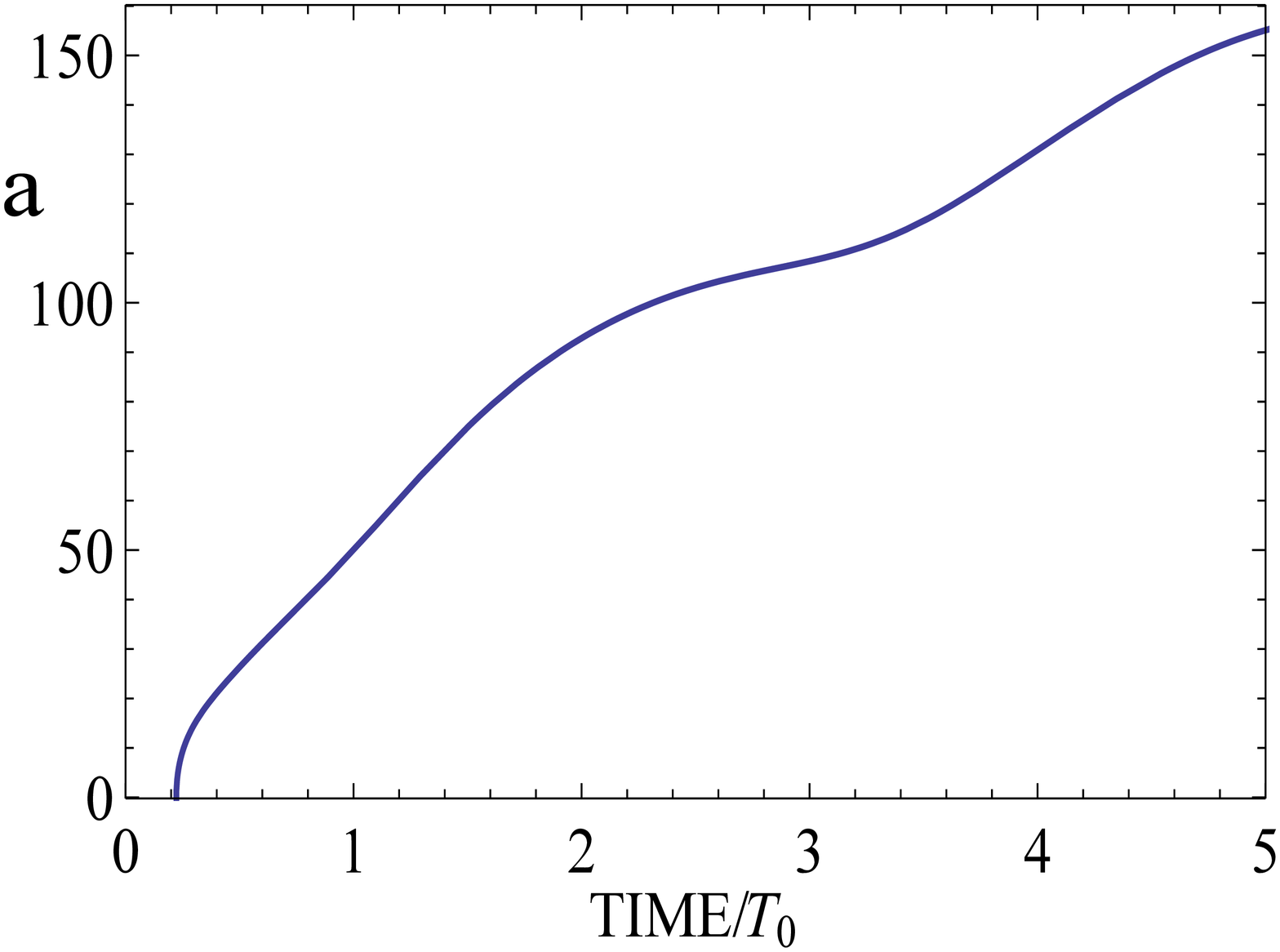}
\includegraphics[width=0.333\textwidth]{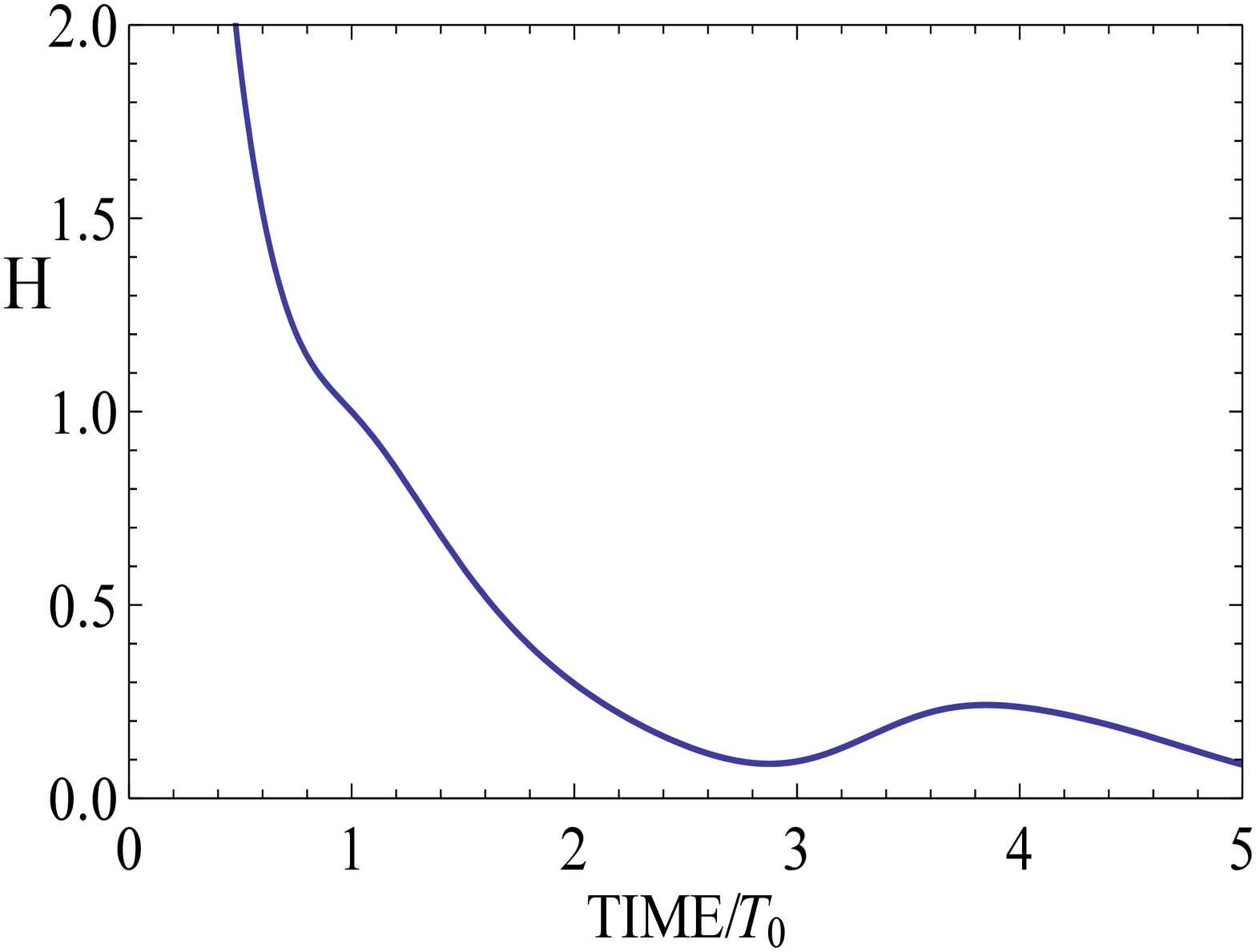}
\includegraphics[width=0.333\textwidth]{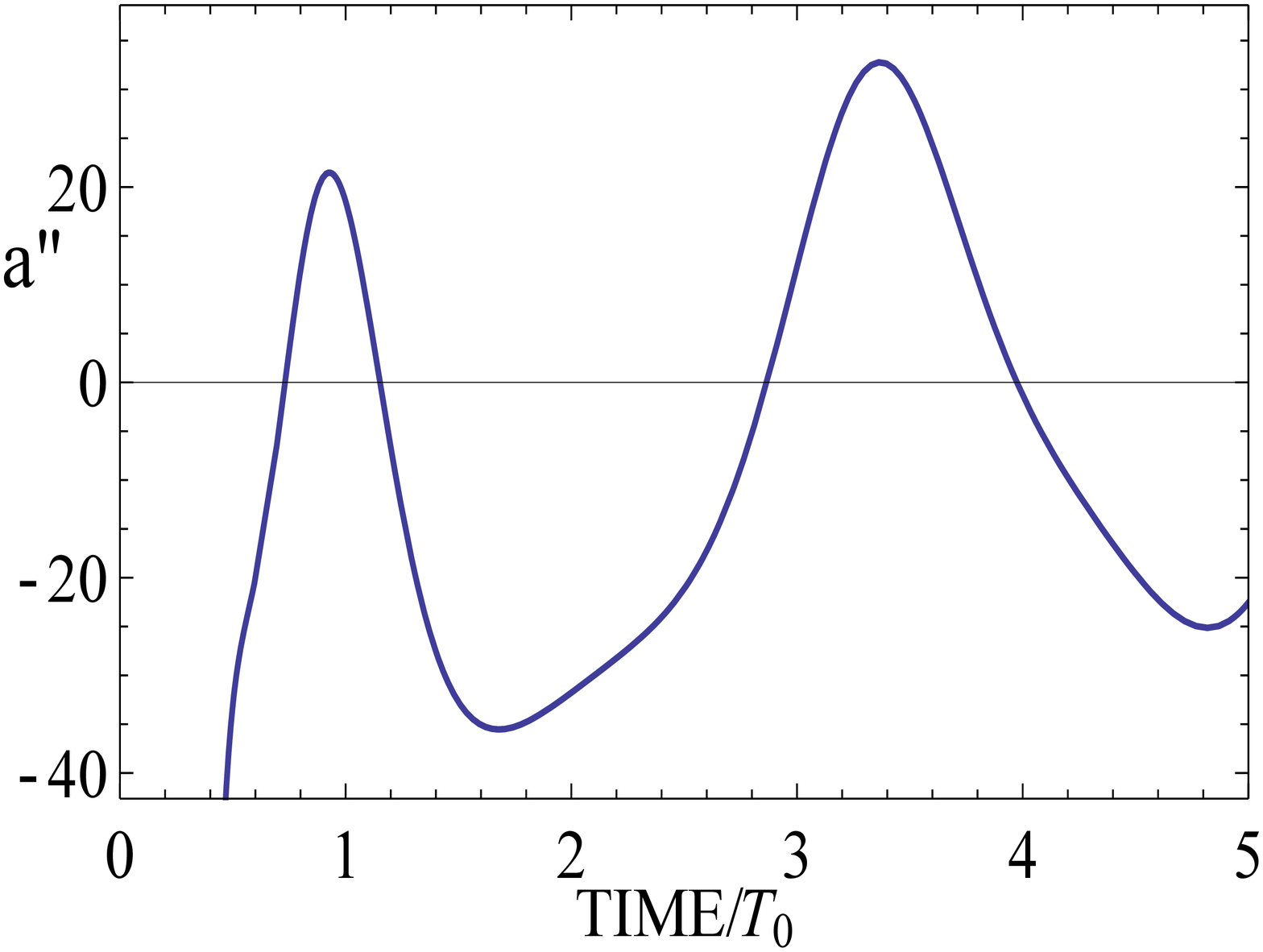}\\
\includegraphics[width=0.333\textwidth]{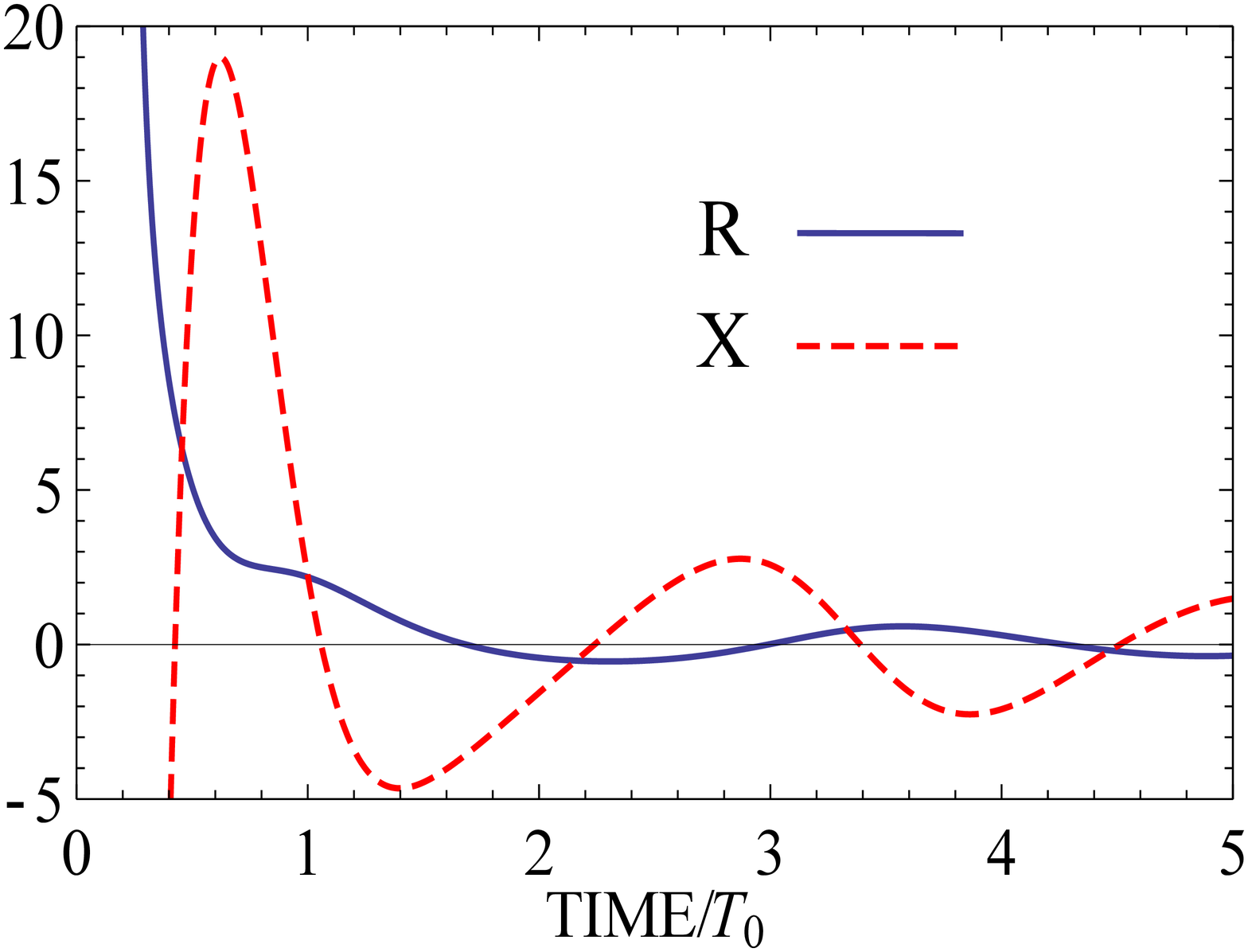}
\includegraphics[width=0.333\textwidth]{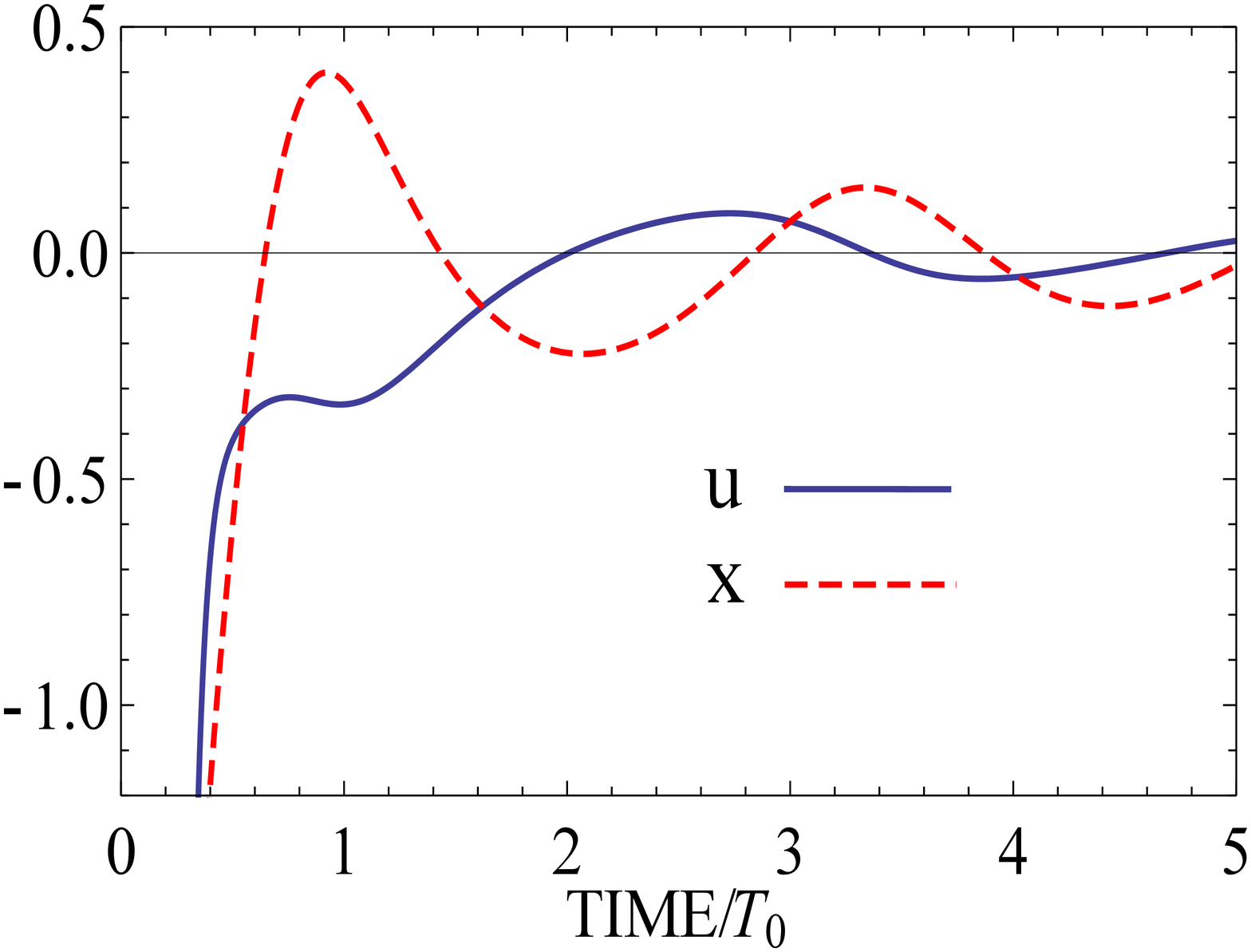}
\includegraphics[width=0.333\textwidth]{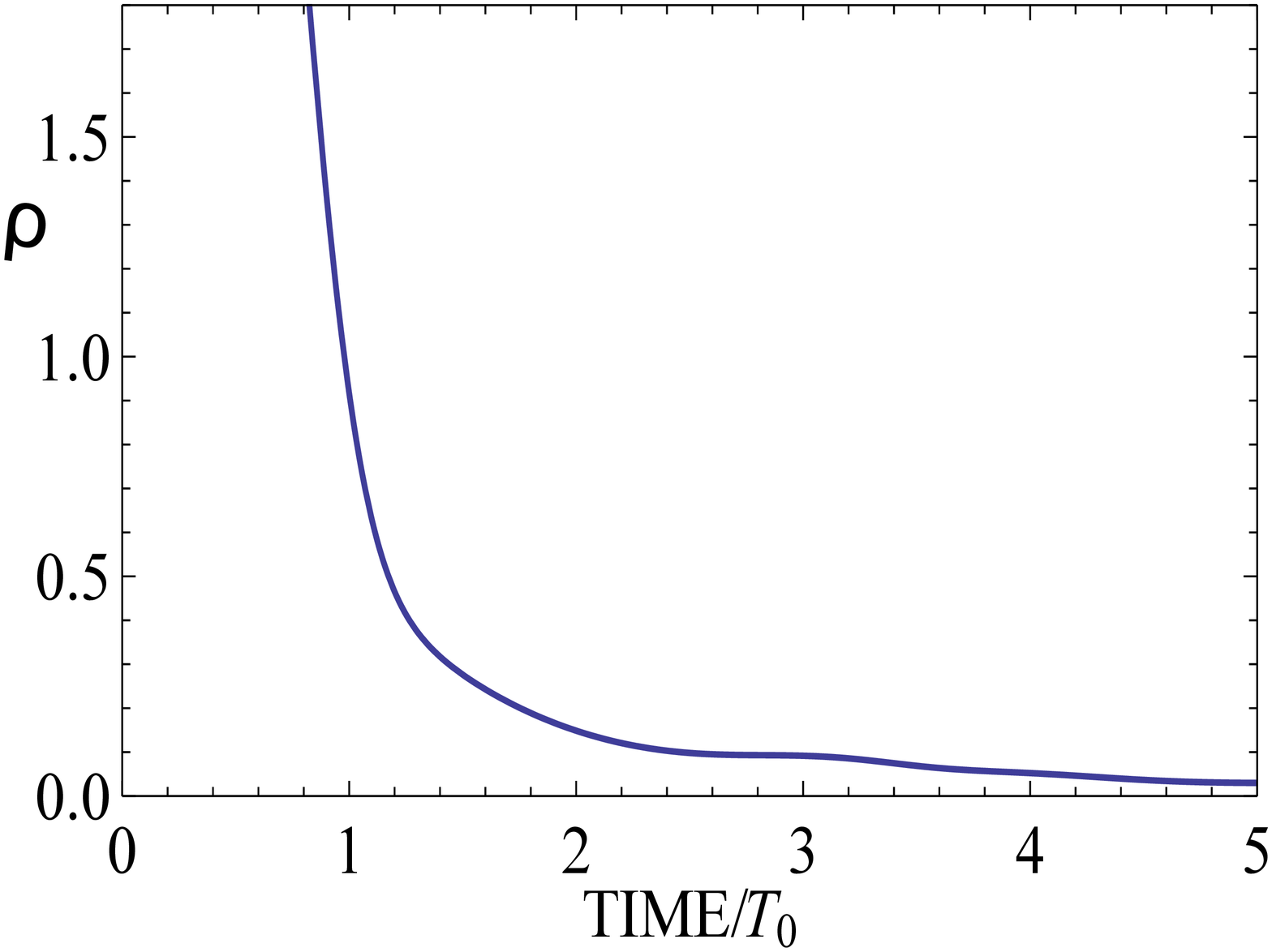}\\
\end{tabular}
\caption{Evolution of the expansion factor a, the Hubble function, $H$,
 the second time derivative of the expansion factor, $\ddot{a}$,
  the scalar curvature and the pseudoscalar curvature, $R$ and $X$,
   the torsion and the axial torsion function, $u$ and $x$ and the energy density, $\rho$.}\label{zA1}
\end{figure}
We present the results of a numerical evolution of
our cosmological model. For all these calculations we take
$\Lambda=0$. We need to look into the scaling features of this model before we
can obtain the sort of evolution results we seek on a cosmological
scale. In terms of fundamental units we can scale the variables and
the parameters as $\kappa=8\pi G=1$. So the variables and the scaled
parameters $w_6$ and $w_3$ become dimensionless (from the Newtonian
limit $a_0=1$). However, as we are interested in the cosmological
scale to see changes of the age of our Universe, let us
introduce a dimensionless constant $T_0$, which represents the
magnitude of the Hubble time ($T_0=H^{-1}_0\doteq4.41504\times
10^{17}$ seconds). With this scaling, all the field equations are kept unchanged while
the period $T\rightarrow T_0T$.

We take the parameters as
\begin{equation}
a_0=1,\quad a_2=0.83,\quad a_3=-0.35,\quad w_{6}=1.1,\quad w_{3}=0.091,\quad \sigma_2=0.12 \quad \hbox{and} \quad \mu_3=-0.18.
\end{equation}
The behavior of the 6 equations has been observed with several
sets of initial values. We plot the typical case in (Fig.~\ref{zA1}):
first, evolution of the expansion factor a, second, the Hubble function, $H$,
third, the second time derivative of the expansion factor, $\ddot{a}$,
fourth, the scalar curvature and the pseudoscalar curvature, $R$ and $X$,
fifth, he torsion and the axial torsion function, $u$ and $x$,
sixth, the energy density, $\rho$.


\section{Concluding discussion}

 We have been investigating the dynamics of the Poincar\'e gauge theory of gravity.
  Recently, the model  with two good propagating modes carrying spin $0^+$, $0^-$
  (referred to as the scalar and pseudoscalar modes) has been extended to
  include pseudoscalar constants that couple the two different parity modes.
  It should be mentioned that here are many interesting physical and mathematical details related to torsion and parity and the type of models considered that we could not discuss here which are covered in  references~\cite{BHN11,SNY08,JCAP09}. Here we have considered the dynamics of this BHN model in the context of
  manifestly homogeneous and isotropic cosmological models.  We found an effective Lagrangian
 and a system of first order equations.  
In these models, at late times the acceleration oscillates.  It can be positive
at the present time.
It should be noted that the $0^+$ torsion does not directly couple to any known form of matter, but it does couple directly to the Hubble expansion, and thus can directly influence the acceleration of the universe.
On the other hand, the $0^-$ couples directly to fundamental fermions; with the newly introduced pseudoscalar coupling constants it too can directly influence the cosmic acceleration.

\section*{Acknowledgments}
This work was supported by the National Science Council of the R.O.C. under the grants NSC-98-2112-M-008-008 and NSC-99-2112-M-008-004 and in part by the National Center of Theoretical Sciences (NCTS).

\section*{References}

\end{document}